\documentclass[11pt,a4paper,notitlepage]{article}

\usepackage{jcappub}

\usepackage{graphicx}
\usepackage{enumitem}
\usepackage{amsmath}
\usepackage{booktabs}
\usepackage{array}
\usepackage{color}
\usepackage{amssymb}
\usepackage{mathrsfs}
\usepackage[T1]{fontenc}
\usepackage[latin1]{inputenc}
\usepackage[table]{xcolor}  
\usepackage{verbatim}
\usepackage{caption}
\usepackage{subcaption}
\usepackage{rotating}
\usepackage{dcolumn}
\usepackage{bm}

\renewcommand{\d}{\mathrm{d}}
\newcommand{\e}[1]{\mathrm{e}^{{#1}}}
\newcommand{\grad}{\nabla}
\newcommand{\vect}[1]{\bm{\mathrm{{#1}}}}
\newcommand{\im}{\mathrm{i}}

\newcommand{\ipleft}{\langle\kern-0.2em\langle}
\newcommand{\ipright}{\rangle\kern-0.2em\rangle}

\DeclareMathOperator{\Lik}{\mathscr{L}}

\newcommand{\Cov}{\hat{\mathcal{C}}}
\newcommand{\Fisher}{\hat{\mathcal{F}}}

\newcommand{\Lag}[1]{\mathcal{L}_{{#1}}}

\newcommand{\cs}{c_{\mathrm{s}}}

\newcommand{\Mpc}{\text{Mpc}}

\renewcommand{\leq}{\leqslant}

\newcommand{\para}[1]{\par\vspace{2mm}\noindent\textbf{{#1.}---}}

\newcolumntype{s}{>{$\displaystyle}l<{$}}
\newcolumntype{t}{>{$\displaystyle}c<{$}}
\newcolumntype{u}{>{$\displaystyle}r<{$}}
\newcolumntype{v}{>{$\displaystyle}m{4cm}<{$}}

\newcolumntype{d}{D{!}{\;\pm\;}{-1}}

\title{Constraining Galileon Inflation}
\author{Donough Regan,$^1$}
\author{Gemma J. Anderson,$^1$}
\author{Matthew Hull$^{1,2}$}
\author{and David Seery$^1$}
\affiliation{\small
$^1$ Astronomy Centre, University of Sussex,
Falmer, Brighton BN1 9QH, UK \\
$^2$ Institute of Cosmology and Gravitation,
University of Portsmouth,
Portsmouth PO1 3FX, UK}
\emailAdd{D.Regan@sussex.ac.uk}
\emailAdd{G.Anderson@sussex.ac.uk}
\emailAdd{Matthew.Hull@port.ac.uk}
\emailAdd{D.Seery@sussex.ac.uk}
\abstract{In this short paper, we present constraints on the
Galileon inflationary model from the CMB bispectrum.
We employ a principal-component analysis of the independent
degrees of freedom constrained by data
and apply this to the WMAP 9-year data
to constrain the free parameters of the model.
A simple Bayesian comparison establishes that
support for the Galileon model
from bispectrum data is at best weak.}
\notoc

\begin{document}	
\maketitle

\section{Introduction}

Recent microwave background data from
the Planck satellite suggest
that the pattern of density fluctuations in our universe is consistent
with a canonical, single-field, slow-roll inflationary model~\cite{Ade:2013ydc}.
To test for deviations from this paradigm we
typically search for signatures in the
the $n$-point functions of the microwave background anisotropies.
At the time of writing, meaningful constraints have been obtained for
the cases $n=2$ and $n=3$---respectively, the power spectrum and
the bispectrum,
corresponding to spectral decompositions of the variance and skewness.

In this paper we focus on searches using the bispectrum, usually
conducted by
comparing fixed `templates' to the data.
This is useful in a discovery phase, where the relevant question
is only whether evidence exists for the amplitude of \emph{some}
template to be inconsistent with zero.
However, because templates do not accurately explore
the range of shapes
produced in a specific model,
it would be more satisfactory
to search for evidence for the model as a whole, rather
than focusing on separate templates.

How should this be done?
A given experiment measures each angular component
of the bispectrum with varying signal-to-noise,
depending on its instrumental characteristics.
Therefore different experiments are sensitive
to differing contributions to the bispectrum.
For any chosen experiment,
a typical model will predict contributions to
which it is highly sensitive
and others to which it is comparatively blind.
We can expect to constrain only those parameters of
a model which contribute to regions of sufficient
sensitivity.
Fitting our models
to these regions simultaneously
gives a balanced picture of the goodness-of-fit
associated with the experiment.
Fitting separate templates may not produce
such a balanced picture if it fails to take
all experimentally-sensitive regions into account.
\newpage
Byun \& Bean have
used this approach to
develop forecasts for a Planck-like experiment~\cite{Byun:2013jba}.
More recently, some of us
applied similar reasoning to the WMAP 9-year dataset
and a very general model described by the effective
theory of single-field inflation~\cite{Anderson:2014mga}.
This theory describes the most general pattern of
fluctuations which can be realized in a Lorentz-invariant
field theory,
assuming Lorentz invariance to be spontaneously broken
by a nearly de Sitter background.
Because it can describe any adiabatic fluctuation
with sufficiently smooth statistical properties
it can be regarded as a weak prior---%
and, as for any prior,
more stringent constraints can be obtained by
strengthening it.
One reason for doing so is to explore
how the interpretation of the data
changes as we vary our
assumptions.
Another is to study how the constraints improve
when we commit to a particular model, rather
than allowing for the most general range of possibilities.

In this paper we focus on a particular prior for
the nonlinear stochastic properties of the inflationary
density perturbation---that it was generated during an
era of `Galileon' inflation~\cite{Burrage:2010cu};
see also Refs.~\cite{DeFelice:2011zh,Kobayashi:2010cm,Kobayashi:2011pc,Kobayashi:2011nu,Mizuno:2010ag}.
Galileons are scalar fields with highly
constrained self-interactions
which contain higher-order time derivatives.
These cancel in the equations of motion~\cite{Nicolis:2008in},
yielding stable second-order field equations.
When quantized this implies that the theory is ghost-free,
and therefore
maintains unitarity and stability.

These stability properties
are preserved by quantum fluctuations
around flat Minkowski space.
At present it is apparently unclear whether the ghost-free
theory can arise as an effective description of
a theory with an ultraviolet completion~\cite{Kaloper:2014vqa},
which would require
the special Galileon self-interactions
to be unaccompanied by other higher-derivative operators
which would generate a ghost.
It is also unknown
whether the ghost-free property survives
on a cosmological background or in the field
created by a heavy source.
But if these possible complications can be evaded
and a field with Galileon-like interactions were
dynamically important during inflation,
then it is possible that their special nature
could leave interesting signatures in the
stochastic properties of the density
perturbation~\cite{Creminelli:2010qf,Burrage:2011hd,Ribeiro:2011ax,Leon:2012mt}.

Our principal result is a constraint on the
importance of the Galileon self-interactions
which would would generate these signatures.
For this purpose the Galileon model is particularly interesting
because it allows just three 3-body interactions
compared to the eleven allowed by the unconstrained
effective field theory.
In Ref.~\cite{Anderson:2014mga} we argued that the WMAP 9-year
dataset is sensitive to three or (at most) four
characteristic contributions to the bispectrum.
Unless we are unlucky and the Galileon 3-body interactions
contribute to these regions in a degenerate way,
we can expect to obtain constraints on all three
couplings.

{More generally, the operators of the Galileon Lagrangian form a subset of the class of single-field effective theories of inflation. The symmetries of the Galileon model impose a relationship between the coefficients of the possible Lagrangian terms. In this paper we exploit this relationship to present the strongest possible constraints on the Galileon paradigm.
}

\para{Notation}%
In~\S\ref{sec:Galileon} we briefly describe the action for
the Galileon inflationary model up to third order,
and in~\S\ref{sec:bisp} we use it to derive the bispectrum.
In~\S\ref{sec:estims} we describe the procedure used to obtain our constraints.
This is a summary of the approach developed in Ref.~\cite{Anderson:2014mga}.
We obtain constraints on the couplings in the Lagrangian for the cases of
one, two or three (the most general possibility) independent third-order couplings. 
Finally, in~\S\ref{sec:bayes} we carry out a Bayesian model comparison
for the various incarnations of the Galileon model.
We give our conclusions in~\S\ref{sec:conclu}.


\section{Overview of Galileon inflation}
\label{sec:Galileon}
The original Galileon model
constructed by Nicolis et al.
was based on a
field-space generalization
of the Galilean shift-symmetry of classical
mechanics,
$\phi \rightarrow \delta_g \phi = \phi + b_\mu x^\mu + c$
where $x^\mu$ is a spacetime coordinate
and $b_\mu$ and $c$ are constants~\cite{Nicolis:2004qq}.
Nicolis et al. worked in flat
spacetime and constructed four
operators, labelled $\Lag{i}$
for $2 \leq i \leq 5$,
which yielded an action satisfying this symmetry%
    \footnote{The $\Lag{i}$ themselves need not be invariant,
    provided that the transformation shifts them only by
    a total derivative which vanishes on integration.}
and produced second-order
equations of motion.
On a curved background it is not possible to retain
both properties~\cite{Deffayet:2009wt,Deffayet:2009mn,deRham:2010eu}.
Insisting on second-order equations of motion and
accepting a break of the shift symmetry
proportional to the background curvature
yields the `covariantized' formulation,
with action
\begin{equation}
    S \supseteq \int \d^4 x \; \sqrt{-g}
    \big[
        c_2 \Lag{2}
        + c_3 \Lag{3}
        + c_4 \Lag{4}
        + c_5 \Lag{5}
    \big] ,
    \label{GalileonLag}
\end{equation}
and $\Lag{i}$ defined by
\begin{subequations}
\begin{align}
    \Lag{2} & = \frac{1}{2} ( \grad \phi )^2 \\
    \Lag{3} & = \frac{1}{\Lambda^3} \Box \phi ( \grad \phi )^2 \\
    \Lag{4} & = \frac{( \grad \phi )^2}{\Lambda^6}
        \big[
            (\Box \phi)^2
            - \grad_\mu \grad_\nu \phi \grad^\mu \grad^\nu \phi - \frac{R}{4} ( \grad \phi )^2
        \big] \\
    \Lag{5} & = \frac{( \grad \phi )^2}{\Lambda^9}
        \big[
            (\Box \phi)^3
            - 3 \Box \phi \grad_\mu \grad_\nu \phi \grad^\mu \grad^\nu \phi
            + 2 \grad_\mu \grad_\nu \phi \grad^\nu \grad^\rho \phi \grad_\rho \grad^\mu \phi
            - 6 G_{\mu\nu} \grad^\mu \grad^\rho \phi \grad^\nu \phi \grad_\rho \phi
        \big] .
\end{align}
\end{subequations}
Here, $G_{\mu\nu}$ is the Einstein tensor,
$R$ denotes the scalar curvature of the background and
$\Lambda$ is a mass scale at which the higher-order
operators $\Lag{3}$, $\Lag{4}$ and $\Lag{5}$ become
comparable to the Gaussian term $\Lag{2}$.

These are not the only
nonlinear
operators which yield an action invariant under the shift
symmetry in flat space---for example,
any function of $\Box \phi$ is automatically
invariant---but they are the only combinations which produce
second-order equations of motion.
Consistency of the model requires that
unwanted combinations such as $(\Box \phi)^2 / M^2$
(which would generally indicate
the presence of a ghost at the scale $M$)
are not generated by renormalization-group running
for any $M$
at which we wish to
trust the predictions of the effective theory.%
    \footnote{It is inconsequential if a ghost is generated
    at scales which are not intended to be described
    by the effective Lagrangian: this happens generically
    in any effective field theory.
    To understand whether the putative ghost really exists
    in the spectrum we would need details of the
    ultraviolet completion.}
This does not happen in the vacuum, where
the $\phi$ fluctuations are massless and
do not generate renormalization-group evolution.
It is not yet known whether a ghost will appear
for non-vacuum field configurations~\cite{Kaloper:2014vqa}.

In this paper we assume that the action~\eqref{GalileonLag}
can be used to describe fluctuations on a quasi-de Sitter background
representing an inflationary phase, and that
there is a regime for which
quantum effects do not cause
ghost modes to become excited.
Working in the `decoupling'
limit where gravitational
degrees of freedom can be ignored,
it was shown in Ref.~\cite{Burrage:2010cu}
that the action for small fluctuations can be written
\begin{equation}
    \label{eq:fluctuation-lagrangian}
    S \supseteq \int \d^4 x \, a^{3}
    \Big[
        \alpha
        \Big( 
            \dot{\pi}^2
            - \frac{\cs^2}{a^{2}} (\partial\pi)^2
        \Big)
        + g_1 \dot{\pi}^3 + \frac{g_{3}}{a^{2}} \dot{\pi} (\partial\pi)^2
        + \frac{g_4}{a^{4}}(\partial\pi)^2\partial^{2}\pi
    \Big] .
\end{equation}
The decoupling limit
applies when the higher-order terms
$\Lag{3}$, $\Lag{4}$ and $\Lag{5}$
are relevant,
making Galileon self-interactions stronger
than gravitational interactions.
But
if pushed too far these self-interactions require
a rapidly evolving field configuration
which risks spoiling the de Sitter background.
A complete understanding of what happens in this regime
would require an analysis of the associated `cosmological'
Vainshtein effect, which has not yet been carried out.
We assume that these complications can be evaded
by having the higher-order terms sufficiently relevant
that they dominate gravitational corrections, but not so
relevant that they destabilize the inflationary era.
The coefficients $\alpha$, $c_s$, $g_2$, $g_3$ and $g_4$
are given by various combinations of $c_i$, $H$, $\dot{\phi}$
and the nonlinear scale $\Lambda$.
For precise formulae, or further discussion
of the role of the nonlinear terms,
we refer to Burrage et al.~\cite{Burrage:2010cu}.
On superhorizon scales the curvature perturbation is given by $\zeta = H\pi$,
and is conserved.

Constraints on this model
take the form of limits on the parameters $c_i$.
Some limits exist based on short-distance
gravitational effects in the late universe;
see, for example, Ref.~\cite{Burrage:2010rs}.
There is no particular requirement for
the Galileon-like fields relevant during the
early and late universe to have the same identity
(although this may be the case in certain models),
so these limits need not apply during inflation.
In this paper we obtain limits on the $c_i$
from the bispectrum of the inflationary density
perturbation
without making use of any late-universe data.

{Note that the fluctuations generated in certain $k$-inflation and Horndeski models may be controlled by the same action~\cite{RenauxPetel:2011sb}. Therefore our results can
equally be interpreted as constraints on these models, although we do not identify them explicitly. Within this large class of theories,
Galileons are algebraically special in that they require only three independent cubic operators, as in Eq.~\eqref{eq:fluctuation-lagrangian}. A generic $k$-inflation or Horndeski model may require up to four independent operators~\cite{RenauxPetel:2011sb,Burrage:2011hd}.}
%
%
%
%
%

\section{Bispectrum Shapes}\label{sec:bisp}
The bispectrum, $B_{\zeta}$, is defined by the three-point
correlation function of the curvature perturbation,
\begin{equation}
    \langle \zeta(\vect{k}_1) \zeta(\vect{k}_2) \zeta(\vect{k}_3) \rangle
    = (2\pi)^3 \delta(\vect{k}_1 + \vect{k}_2 + \vect{k}_3) B_\zeta(k_1, k_2, k_3) .
\end{equation}

\para{Wavefunctions}%
A complication of the most general models studied in Ref.~\cite{Anderson:2014mga}
is that fourth-derivative operators can appear in the quadratic term,
leading to a very complex form for the elementary functions.
While a solution can be found in closed form, the subsequent vertex integrations
appearing in the Feynmann rules cannot be performed analytically.
The Galileon fluctuation Lagrangian~\eqref{GalileonLag}
belongs to the class of models considered
Ref.~\cite{Anderson:2014mga} but does not possess the problematic
fourth-derivative operators.
Neglecting slow-roll corrections the elementary wavefunction is
\begin{equation}
    \label{eq:leadingwave}
    u(k,\tau)
    =
        \frac{\im H}{2\sqrt{\alpha}}
        \frac{1}{(k c_s)^{3/2}} (1 - \im k c_s \tau) \e{\im k c_s \tau} .
\end{equation}
On superhorizon scales where $| k c_s \tau | \ll 1$
the power spectrum can be written
\begin{equation}
    P_{\zeta}(k)=\frac{H^4}{4 \alpha  c_s^3}\frac{1}{k^3} .
\end{equation}


\para{Momentum dependence}%
The necessary calculations were described by
Mizuno \& Koyama~\cite{Mizuno:2010ag},
and given to next-order in slow-roll parameters in
Refs.~\cite{Burrage:2010cu,Burrage:2011hd}.
Similar calculations were performed by
Kobayashi, Yamaguchi and Yokoyama~\cite{Kobayashi:2011pc}.
There is one contribution to the bispectrum
from each cubic operator in~\eqref{eq:fluctuation-lagrangian},
which we write in the form
\begin{equation}
    \label{eq:zeta-bispectrum}
    B_\zeta(k_1, k_2, k_3) = \frac{3}{5} 
        \sum_{\alpha=1}^3 \lambda_\alpha B_{\alpha} (k_1, k_2, k_3) .
\end{equation}
The $B_\alpha$ are normalized so that
$B_{\alpha}(k,k,k)/6{P}_\zeta(k)^2 = 1$
at the equilateral point.
In terms of the couplings $g_\alpha$ in the fluctuation Lagrangian
these means that that $\lambda_\alpha$
correspond to
\begin{equation}
    \label{eq:lambdaversusg}
    \lambda_1 =
        \frac{5}{81}\frac{g_1}{\alpha}
    \qquad
    \lambda_2 =
        -\frac{85}{324}\frac{g_3}{ c_s^2 \alpha}
    \qquad
    \lambda_3 =
        -\frac{65}{162}\frac{g_4 H}{c_s^4 \alpha} .
\end{equation}
Focusing only on the momentum dependence
(prefactors can be inferred from the normalization convention if required),
the bispectra can be written
\begin{subequations}
\begin{align}
    \label{eq:b1}
    B_{1}(k_1,k_2,k_3) & \sim
        \frac{1}{\prod\limits_{i} k_i}\frac{1}{k_{t}^3}
    \\
    \label{eq:b2}
    B_{2}(k_1,k_2,k_3) & \sim
        \frac{1}{\prod\limits_{i}k_{i}^{3}} k_{1}^{2}
        (\vect{k}_{2} \cdot\vect{k}_{3})
        \left(
            \frac{1}{k_{t}}
            + \frac{k_{2}+k_{3}}{k_{t}^2}
            + \frac{2k_{2}k_{3}}{k_{t}^3}
        \right)
        + 1 \! \rightarrow \! 2
        + 1 \! \rightarrow \! 3
    \\
    \label{eq:b3}
    B_{3}(k_1,k_2,k_3) & \sim
        \frac{1}{\prod\limits_{i}k_{i}^{3}}\,k_{1}^{2}
        (\vect{k}_{2}\cdot\vect{k}_{3})
        \left(
            \frac{1}{k_{t}}
            + \frac{K^2}{k_{t}^3}
            + \frac{3k_1 k_2 k_3}{k_{t}^4}
        \right)
        + 1 \!\rightarrow \! 2
        + 1 \! \rightarrow \! 3 ,
\end{align}
\end{subequations}
where $k_t = k_1 + k_2 + k_3$,
and
$K^2 = k_1k_2 + k_1k_3 + k_2k_3$.
 
\section{Estimating Galileon Parameters}\label{sec:estims}
We now aim to estimate the $\lambda_\alpha$ using CMB data.
Given a model, and therefore knowledge of the parameters
$\alpha$ and $c_s$, this enables the $g_\alpha$ to be
determined.
Given knowledge of the background field configuration this
enables constraints to be placed on the $c_i$.

\para{Estimation methodology}%
Our methodology for estimating the $\lambda_\alpha$
was described in Ref.~\cite{Anderson:2014mga}.
Following Fergusson, Liguori \& Shellard we
write each of~\eqref{eq:b1}--\eqref{eq:b3}
as a sum over some basis $B_n$, giving
$B_\alpha = \sum_n \alpha_\alpha^n B_n$~\cite{Fergusson:2009nv}.
We extract multipole coefficients for the
temperature anisotropy $\Delta T$
using $\Delta T(\hat{\vect{n}})/T = \sum_{\ell m} a_{\ell m} Y_{\ell m}(\hat{\vect{n}})$
and define the angular bispectrum
$b_{\ell_1 \ell_2 \ell_3}$
to satisfy
$\langle a_{\ell_1 m_1} a_{\ell_2 m_2} a_{\ell_3 m_3} \rangle
= b_{\ell_1 \ell_2 \ell_3} \mathcal{G}^{\ell_1 \ell_2 \ell_3}_{m_1 m_2 m_3}$,
where $\mathcal{G}^{\ell_1 \ell_2 \ell_3}_{m_1 m_2 m_3}$ is the 
Gaunt integral.
After using the primordial perturbation $\zeta$ to seed
fluctuations in the radiation era,
and accounting for radiative transfer
and projection onto the sky,
each $B_n$ will yield some angular bispectrum
$b^n_{\ell_1 \ell_2 \ell_3}$.
We introduce a further basis $b^A_{\ell_1 \ell_2 \ell_3}$
and write
$b^n_{\ell_1 \ell_2 \ell_3} = {\Gamma_A}^n b^A_{\ell_1 \ell_2 \ell_3}$.
Then, given a choice $\lambda_\alpha$, it follows that the observable
angular bispectrum can be written
\begin{equation}
    b_{\ell_1 \ell_2 \ell_3} = \sum_A \beta_A b^A_{\ell_1 \ell_2 \ell_3} ,
\end{equation}
where the coefficients $\beta_A$ are defined by
\begin{equation}
    \beta_A \equiv \lambda_\alpha \alpha^\alpha_n {\Gamma_A}^n .
\end{equation}
The advantage of this basis decomposition
is that the transfer matrix ${\Gamma_A}^n$ can be computed
relatively easily \cite{Fergusson:2010gn}.
It encodes details of the cosmology, together with
the processes of radiative transfer
which connect primordial times to observation.

A given microwave background experiment makes measurements
of the $\beta_A$ associated with our last-scattering surface.
We write these estimates $\hat{\beta}_A$.
Assuming Gaussian experimental errors, and
given our prior, the likelihood of an experiment returning
some particular set of values can be written
\begin{equation}
    \label{eq:likelihood}
    \Lik( \hat{\beta}_A | \lambda_\alpha )
    =
    \frac{1}{\sqrt{2\pi \det \Cov}}
    \exp\bigg(
        - \frac{1}{2} \sum_{A, B} (\Cov^{-1})^{AB}
            \Delta \hat{\beta}_A
            \Delta \hat{\beta}_B
    \bigg) ,
\end{equation}
where $\Delta \hat{\beta}_A \equiv \hat{\beta}_A - \beta_A$
and
the covariance matrix is defined by
$C_{AB} = \langle \Delta \hat{\beta}_A \Delta \hat{\beta}_B \rangle$.
We estimate it using Gaussian simulations,
accounting for
realistic WMAP beam and noise properties
and the effects of masking.
For all quantitative details we refer to Ref.~\cite{Anderson:2014mga}.

Eq.~\eqref{eq:likelihood}
yields a
maximum likelihood estimator
for each parameter $\lambda_\alpha$ corresponding to
\begin{equation}
    \hat{\lambda}_\alpha
    =
    \sum_\beta \hat{b}^\beta (\Fisher^{-1})_{\beta\alpha} ,
\end{equation}
where
$\hat{b}^\alpha \equiv \sum_{A, B, n} \hat{\beta}_A (\Cov^{-1})^{AB}
\alpha^\alpha_n {\Gamma_B}^n$
and the Fisher matrix $\vect{\Fisher}$ satisfies
\begin{equation}
    \Fisher^{\alpha\beta} =\sum_{A,B,m,n}
        {\Gamma_A}^m \alpha^\alpha_m
        (\Cov^{-1})^{AB}
        \alpha^\beta_n
        {\Gamma_B}^n .
\end{equation}

\para{Application to 9-year WMAP data}%
We use the WMAP 9-year dataset to estimate the
amplitudes $\hat{\beta}_A$~\cite{Hinshaw:2012aka,Bennett:2012zja},
and use these to constrain
subcases of the fluctuation Lagrangian~\eqref{eq:fluctuation-lagrangian}.
The most general case includes all three operators
$g_1$, $g_3$ and $g_4$
and yields the weakest constraints.
This would be expected where correlations among
the operators exist in the regions to which WMAP is most sensitive.
In this case rather more is true:
Renaux-Petel pointed out~\cite{RenauxPetel:2011sb,Renaux-Petel:2013ppa}
that there is an approximate degeneracy
(spoilt by boundary terms which become irrelevant at late times)
which allows the $g_4$ contribution to be absorbed into renormalizations
of the other couplings,
\begin{align}
    \label{eq:g13eqn}
    g_1\rightarrow g_1'=g_1+g_4 H/c_s^4 ,
    \qquad
    g_3\rightarrow g_3'=g_3+2 g_4 H/c_s^2.
\end{align}
We can leave $g_4$ in the analysis, accounting
for the correlation in shape, or
eliminate $g_4$ using~\eqref{eq:g13eqn}
at the outset.
In what follows we will give constraints for both choices.
Finally, we consider the most restrictive subcase in which
only one parameter is allowed to be nonzero.
This corresponds most directly with the standard approach
of fitting individual templates to the data.
It gives optimistic constraints unless
we are prepared to commit to a scenario in which
two operators are subdominant compared to the third.

\para{Case 1: General scenario (three free parameters)}%
Using the relationship between the parameters
$\lambda_\alpha$ and the coefficients of the Lagrangian
given in Eq.~\eqref{eq:lambdaversusg}, we find
\begin{center}
    \small
	\heavyrulewidth=.08em
	\lightrulewidth=.05em
	\cmidrulewidth=.03em
	\belowrulesep=.65ex
	\belowbottomsep=0pt
	\aboverulesep=.4ex
	\abovetopsep=0pt
	\cmidrulesep=\doublerulesep
	\cmidrulekern=.5em
	\defaultaddspace=.5em
	\renewcommand{\arraystretch}{1.5}
        
    \rowcolors{2}{gray!25}{white}

    \begin{tabular}{sd}
        \toprule
       \text{Variable}   & \multicolumn{1}{c}{Estimate} \\
        {\hat{g}_1}/{\alpha} & (4.21 ! 3.96) \times 10^5 \\
       {\hat{g}_3}/{c_s^2\alpha} & (4.18 !  4.03) \times 10^5 \\
       {\hat{g}_4 H}/{c_s^4\alpha}& (-2.09 ! 2.04) \times 10^5  \\
        \bottomrule

    \end{tabular}
\end{center}
The quoted uncertainties represent $1\sigma$
errors bars, with marginalization over the other two parameters.
Each constraint is consistent with zero to within $1\sigma$.
Note that the uncertainties are rather large, due to exploration
of the entire parameter space.

\para{Case 2: Two free parameters}%
It was explained above that the three-parameter case
is perhaps too pessimistic, because of correlations between
the bispectra produced by the three cubic operators.
In this section we obtain constraints on the subcase where
two couplings are allowed to vary with the third fixed at zero.
In the case where $g_4$ is held fixed Eq.~\eqref{eq:g13eqn}
can be used to map these constraints to the Lagrangian obtained
by elimination of the $g_4$ term.
The results are
\begin{center}
    \small
	\heavyrulewidth=.08em
	\lightrulewidth=.05em
	\cmidrulewidth=.03em
	\belowrulesep=.65ex
	\belowbottomsep=0pt
	\aboverulesep=.4ex
	\abovetopsep=0pt
	\cmidrulesep=\doublerulesep
	\cmidrulekern=.5em
	\defaultaddspace=.5em
	\renewcommand{\arraystretch}{1.5}
        
    \rowcolors{2}{gray!25}{white}

    \begin{tabular}{sdd}
        \toprule
        \text{Fixed parameter}     &\text{Variable}    & \multicolumn{1}{c}{Estimate} \\
        {{g}_1}/{\alpha} & {\hat{g}_3}/{c_s^2\alpha} &-11,000   !  6,770   \\
        \rowcolor{white}
              &   {\hat{g}_4 H}/{c_s^4\alpha} &    7,760!  4,870 \\
                      \rowcolor{gray!25}
        {{g}_3}/{c_s^2\alpha} & {\hat{g}_1}/{\alpha} &  11,000 !  6,660  \\
        \rowcolor{gray!25}
              &   {\hat{g}_4 H}/{c_s^4\alpha} &  3,380 !  2,240   \\
        {{g}_4 H}/{c_s^4\alpha} & {\hat{g}_1}/{\alpha} &  15,300 !  9,470  \\
        \rowcolor{white}
              &   {\hat{g}_3}/{c_s^2\alpha} &    2,870 !  1,900\\
        \bottomrule

    \end{tabular}
\end{center}
Each constraint is consistent with zero
to within $\sim1.5\sigma$,
matching our expectations
from constraints on individual templates using
the 9-year WMAP dataset~\cite{Bennett:2012zja}.
Similar results have been reported from the Planck \cite{Ade:2013ydc} 2013 data release.
The formalism used here ensures that the entire available parameter space
is explored, rather than inferring these constraints from the overlap with a selection of templates.

\para{Case 3: One free parameter}%
Finally, we consider the constraints where only one
parameter is allowed to vary. We obtain
\begin{align}
    {\hat{g}_1}/{\alpha}=1120\pm 1280 , \qquad
    {\hat{g}_3}/{c_s^2\alpha}=-260\pm 390 , \qquad
    {\hat{g}_4 H}/{c_s^4\alpha}=-160\pm 280 .
\end{align}
We may also obtain a constraint from the amplitude of the
power spectrum, which gives $H^4/(\alpha c_s^3)=(190\pm 8 )\times 10^{-9}$
at the pivot scale $k = 0.002 \Mpc^{-1}$.
Together this gives $4$ constraints for $6$ parameters,
$\{H,\alpha,c_s,g_2,g_3,g_4\}$. Breaking the degeneracy
would
require constraints on the scalar tilt, $n_s$ or the tensor to scalar ratio, $r$.

\section{Bayesian Model Comparison}\label{sec:bayes}
While our results in the previous section are useful
in determining best-fit values for the parameters,
we wish also to perform a model comparison. One method with which
to quantify to evidence for or against a model is
through calculation of the `Bayes' factor. In Ref.~\cite{Anderson:2014mga}
this was first applied to the comparison of non-Gaussian models.
We briefly recapitulate the description here.  

Given a data set $D$ and a pair of models $M_1$ and $M_2$,
with respective parameter sets $\{\lambda_1\}$ and $\{\lambda_2\}$,
the Bayes factor is defined as the ratio of the likelihoods of the respective models,
\begin{align}\label{eq:Bayesfac}
    K_{12}
        = \frac{P(D|M_1)}{P(D|M_2)}
        =
        \frac{\int P(D|\{\lambda_1\}, M_1) P(\{\lambda_1\}|M_1)\,\d\lambda_1}
            {\int P(D|\{\lambda_2\}, M_2) P(\{\lambda_2\}|M_2)\,\d\lambda_2} .
\end{align}
It should be noted that the integrals here are over
the entire parameter space of each model.
These may be of different dimensionalities.
The prior probabilities $P(\{\lambda_i\}|M_i)$ represent the
probability that a particular parameter choice occurs.
To determine our priors we use the requirement that the bispectrum
generated by each operator must not dominate the power spectrum,
and therefore that each parameter is constrained by $|\lambda_i|\lesssim 10^4$.
However, given that we have no reason%
    \footnote{We cannot use constraints from CMB experiments to choose our prior,
    because we are using the CMB as our dataset.}
to prefer any scale we choose a Jeffries prior with
$P(\lambda_i)\propto 1/|\lambda_i|$, with $|\lambda_i|\in[1,10^4]$.
The cutoff is chosen to avoid a divergence at $\lambda_i=0$,
and
our results show little dependence on its precise value.
The Bayes factor does not become independent of the prior, so to
study its dependence for different choices we also
compute values for a flat prior in the range $[-1,1]$.
The probability $P(D|\{\lambda_1\}, M_1)$ represents
the likelihood for a particular choice of parameters $\{\lambda_i\}$
and may be computed using Eq.~\eqref{eq:likelihood}. 

We interpret our results using the Kass \& Raftery scale \cite{doi:10.1080/01621459.1995.10476572}.
In this scheme, $|\ln K|$ in the range $(0,1)$ is `indecisive',
in the range $(1,3)$ represents `evidence in favour of $M_1$',
in the range $(3,5)$ represents `strong evidence in favour of $M_1$',
and larger values are `decisive'.
A similar scale applies to $K^{-1}$ with $M_2$ substituted for $M_1$.
\paragraph{Results}
\begin{itemize}
\item Comparing the Gaussian model (i.e. with all $\lambda_i=0$) with the case of just one non-zero parameter, we find the Bayes factor is given by $\ln K \approx 0.7$, such that the data is indecisive in distinguishing these scenarios.
\item Next we compare the Gaussian model and the case of two free parameters, with the result that $\ln K \approx 1.3$.
This indicates (weak) evidence against the Galileon model with two free parameters, and is mainly due to an Ockham penalty which disfavours addition of extra parameters without sufficient support from the data. Adding a further parameter and comparing the Gaussian model to the most general Galileon model with three free parameters gives a Bayes factor $\ln K \approx 2$.
In this case there is an even stronger preference for the simpler description.
\item Comparing the single free parameter case with the two parameter and three parameter cases giving $\ln K\approx 0.6$ and $\ln K\approx 1.3$, respectively. The data is indecisive in the former case, but shows preference for the single parameter case in the latter.
\end{itemize}
In summary, the data shows little power to discriminate between the Gaussian model
and a Galileon model with one extra free parameter.
However, for models with two or more extra parameters the WMAP 9-year data exhibits
a weak preference for the simpler description.

\section{Conclusions}\label{sec:conclu}
In this paper we have utilised the formalism developed in Ref.~\cite{Anderson:2014mga}
to constrain the Galileon inflationary model using the bispectrum.
Our constraints show that the couplings of the cubic terms in the fluctuation Lagrangian
are consistent with zero to within $1.5\sigma$.
We have separately considered the cases of one, two and three free parameters in the fluctuation Lagrangian.

The formalism can be used to carry out a Bayesian model comparison. This establishes that the data weakly disfavours
models requiring two extra free parameters, but is inconclusive between a Gaussian model and the case of a
Galileon model with a single extra coupling.
It is possible that carrying out the analysis using Planck data~\cite{Ade:2013ydc} may lead to a stronger result.  

\section*{Acknowledgements}
DR and DS acknowledge support from the Science and Technology
Facilities Council [grant number ST/L000652/1].
GA is supported by STFC grant ST/L000652/1. MH is supported by an STFC research studentship. 
DS also acknowledges support from the Leverhulme Trust. Some numerical presented in this paper were obtained using the COSMOS supercomputer, which is funded by STFC, HEFCE and SGI. Other numerical computations were carried out on the Sciama High Performance Compute (HPC) cluster which is supported by the ICG, SEPNet and the University of Portsmouth.
The research leading to these results has received funding from
the European Research Council under the European Union's
Seventh Framework Programme (FP/2007--2013) / ERC Grant
Agreement No. [308082].

\bibliography{bibli}

\providecommand{\href}[2]{#2}\begingroup\raggedright\begin{thebibliography}{10}

\bibitem{Ade:2013ydc}
{\bf Planck Collaboration} Collaboration, P.~Ade {\em et~al.}, {\it {Planck
  2013 Results. XXIV. Constraints on primordial non-Gaussianity}},
  \href{http://arxiv.org/abs/1303.5084}{{\tt arXiv:1303.5084}}.

\bibitem{Byun:2013jba}
J.~Byun and R.~Bean, {\it {Non-Gaussian Shape Recognition}},  {\em JCAP} {\bf
  1309} (2013) 026, [\href{http://arxiv.org/abs/1303.3050}{{\tt
  arXiv:1303.3050}}].

\bibitem{Anderson:2014mga}
G.~J. Anderson, D.~Regan, and D.~Seery, {\it {Optimal bispectrum constraints on
  single-field models of inflation}},  {\em JCAP} {\bf 1407} (2014) 017,
  [\href{http://arxiv.org/abs/1403.3403}{{\tt arXiv:1403.3403}}].

\bibitem{Burrage:2010cu}
C.~Burrage, C.~de~Rham, D.~Seery, and A.~J. Tolley, {\it {Galileon inflation}},
   {\em JCAP} {\bf 1101} (2011) 014,
  [\href{http://arxiv.org/abs/1009.2497}{{\tt arXiv:1009.2497}}].

\bibitem{DeFelice:2011zh}
A.~De~Felice and S.~Tsujikawa, {\it {Primordial non-Gaussianities in general
  modified gravitational models of inflation}},  {\em JCAP} {\bf 1104} (2011)
  029, [\href{http://arxiv.org/abs/1103.1172}{{\tt arXiv:1103.1172}}].

\bibitem{Kobayashi:2010cm}
T.~Kobayashi, M.~Yamaguchi, and J.~Yokoyama, {\it {G-inflation: Inflation
  driven by the Galileon field}},  {\em Phys.Rev.Lett.} {\bf 105} (2010)
  231302, [\href{http://arxiv.org/abs/1008.0603}{{\tt arXiv:1008.0603}}].

\bibitem{Kobayashi:2011pc}
T.~Kobayashi, M.~Yamaguchi, and J.~Yokoyama, {\it {Primordial non-Gaussianity
  from G-inflation}},  {\em Phys.Rev.} {\bf D83} (2011) 103524,
  [\href{http://arxiv.org/abs/1103.1740}{{\tt arXiv:1103.1740}}].

\bibitem{Kobayashi:2011nu}
T.~Kobayashi, M.~Yamaguchi, and J.~Yokoyama, {\it {Generalized G-inflation:
  Inflation with the most general second-order field equations}},  {\em
  Prog.Theor.Phys.} {\bf 126} (2011) 511--529,
  [\href{http://arxiv.org/abs/1105.5723}{{\tt arXiv:1105.5723}}].

\bibitem{Mizuno:2010ag}
S.~Mizuno and K.~Koyama, {\it {Primordial non-Gaussianity from the DBI
  Galileons}},  {\em Phys.Rev.} {\bf D82} (2010) 103518,
  [\href{http://arxiv.org/abs/1009.0677}{{\tt arXiv:1009.0677}}].

\bibitem{Nicolis:2008in}
A.~Nicolis, R.~Rattazzi, and E.~Trincherini, {\it {The Galileon as a local
  modification of gravity}},  {\em Phys.Rev.} {\bf D79} (2009) 064036,
  [\href{http://arxiv.org/abs/0811.2197}{{\tt arXiv:0811.2197}}].

\bibitem{Kaloper:2014vqa}
N.~Kaloper, A.~Padilla, P.~Saffin, and D.~Stefanyszyn, {\it {Unitarity and the
  Vainshtein Mechanism}},  \href{http://arxiv.org/abs/1409.3243}{{\tt
  arXiv:1409.3243}}.

\bibitem{Creminelli:2010qf}
P.~Creminelli, G.~D'Amico, M.~Musso, J.~Norena, and E.~Trincherini, {\it
  {Galilean symmetry in the effective theory of inflation: new shapes of
  non-Gaussianity}},  {\em JCAP} {\bf 1102} (2011) 006,
  [\href{http://arxiv.org/abs/1011.3004}{{\tt arXiv:1011.3004}}].

\bibitem{Burrage:2011hd}
C.~Burrage, R.~H. Ribeiro, and D.~Seery, {\it {Large slow-roll corrections to
  the bispectrum of noncanonical inflation}},  {\em JCAP} {\bf 1107} (2011)
  032, [\href{http://arxiv.org/abs/1103.4126}{{\tt arXiv:1103.4126}}].

\bibitem{Ribeiro:2011ax}
R.~H. Ribeiro and D.~Seery, {\it {Decoding the bispectrum of single-field
  inflation}},  {\em JCAP} {\bf 1110} (2011) 027,
  [\href{http://arxiv.org/abs/1108.3839}{{\tt arXiv:1108.3839}}].

\bibitem{Leon:2012mt}
G.~Leon and E.~N. Saridakis, {\it {Dynamical analysis of generalized Galileon
  cosmology}},  {\em JCAP} {\bf 1303} (2013) 025,
  [\href{http://arxiv.org/abs/1211.3088}{{\tt arXiv:1211.3088}}].

\bibitem{1974IJTP...10..363H}
G.~W. {Horndeski}, {\it {Second-Order Scalar-Tensor Field Equations in a
  Four-Dimensional Space}},  {\em International Journal of Theoretical Physics}
  {\bf 10} (Sept., 1974) 363--384.

\bibitem{Deffayet:2011gz}
C.~Deffayet, X.~Gao, D.~Steer, and G.~Zahariade, {\it {From k-essence to
  generalised Galileons}},  {\em Phys.Rev.} {\bf D84} (2011) 064039,
  [\href{http://arxiv.org/abs/1103.3260}{{\tt arXiv:1103.3260}}].

\bibitem{Nicolis:2004qq}
A.~Nicolis and R.~Rattazzi, {\it {Classical and quantum consistency of the DGP
  model}},  {\em JHEP} {\bf 0406} (2004) 059,
  [\href{http://arxiv.org/abs/hep-th/0404159}{{\tt hep-th/0404159}}].

\bibitem{Deffayet:2009wt}
C.~Deffayet, G.~Esposito-Farese, and A.~Vikman, {\it {Covariant Galileon}},
  {\em Phys.Rev.} {\bf D79} (2009) 084003,
  [\href{http://arxiv.org/abs/0901.1314}{{\tt arXiv:0901.1314}}].

\bibitem{Deffayet:2009mn}
C.~Deffayet, S.~Deser, and G.~Esposito-Farese, {\it {Generalized Galileons: All
  scalar models whose curved background extensions maintain second-order field
  equations and stress-tensors}},  {\em Phys.Rev.} {\bf D80} (2009) 064015,
  [\href{http://arxiv.org/abs/0906.1967}{{\tt arXiv:0906.1967}}].

\bibitem{deRham:2010eu}
C.~de~Rham and A.~J. Tolley, {\it {DBI and the Galileon reunited}},  {\em JCAP}
  {\bf 1005} (2010) 015, [\href{http://arxiv.org/abs/1003.5917}{{\tt
  arXiv:1003.5917}}].

\bibitem{Burrage:2010rs}
C.~Burrage and D.~Seery, {\it {Revisiting fifth forces in the Galileon model}},
   {\em JCAP} {\bf 1008} (2010) 011,
  [\href{http://arxiv.org/abs/1005.1927}{{\tt arXiv:1005.1927}}].

\bibitem{Fergusson:2009nv}
J.~Fergusson, M.~Liguori, and E.~Shellard, {\it {General CMB and Primordial
  Bispectrum Estimation I: Mode Expansion, Map-Making and Measures of
  $f_{NL}$}},  {\em Phys.Rev.} {\bf D82} (2010) 023502,
  [\href{http://arxiv.org/abs/0912.5516}{{\tt arXiv:0912.5516}}].

\bibitem{Fergusson:2010gn}
J.~Fergusson, D.~Regan, and E.~Shellard, {\it {Optimal Trispectrum Estimators
  and WMAP Constraints}},  \href{http://arxiv.org/abs/1012.6039}{{\tt
  arXiv:1012.6039}}.

\bibitem{Hinshaw:2012aka}
{\bf WMAP Collaboration} Collaboration, G.~Hinshaw {\em et~al.}, {\it
  {Nine-Year Wilkinson Microwave Anisotropy Probe (WMAP) Observations:
  Cosmological Parameter Results}},  \href{http://arxiv.org/abs/1212.5226}{{\tt
  arXiv:1212.5226}}.

\bibitem{Bennett:2012zja}
{\bf WMAP} Collaboration, C.~Bennett {\em et~al.}, {\it {Nine-Year Wilkinson
  Microwave Anisotropy Probe (WMAP) Observations: Final Maps and Results}},
  {\em Astrophys.J.Suppl.} {\bf 208} (2013) 20,
  [\href{http://arxiv.org/abs/1212.5225}{{\tt arXiv:1212.5225}}].

\bibitem{RenauxPetel:2011sb}
S.~Renaux-Petel, {\it {On the redundancy of operators and the bispectrum in the
  most general second-order scalar-tensor theory}},  {\em JCAP} {\bf 1202}
  (2012) 020, [\href{http://arxiv.org/abs/1107.5020}{{\tt arXiv:1107.5020}}].

\bibitem{Renaux-Petel:2013ppa}
S.~Renaux-Petel, {\it {DBI Galileon in the Effective Field Theory of Inflation:
  Orthogonal non-Gaussianities and constraints from the Trispectrum}},  {\em
  JCAP} {\bf 1308} (2013) 017, [\href{http://arxiv.org/abs/1303.2618}{{\tt
  arXiv:1303.2618}}].

\bibitem{doi:10.1080/01621459.1995.10476572}
R.~E. Kass and A.~E. Raftery, {\it Bayes factors},  {\em Journal of the
  American Statistical Association} {\bf 90} (1995), no.~430 773--795,
  [\href{http://arxiv.org/abs/http://amstat.tandfonline.com/doi/pdf/10.1080/01621459.1995.10476572}{{\tt
  http://amstat.tandfonline.com/doi/pdf/10.1080/01621459.1995.10476572}}].

\end{thebibliography}\endgroup
\end{document}